# Another Multiparametric Way In Planning Of Experiments For O/W Emulsion Design

**Guyader, Natacha**[1*]; Michiel, Magalie[1]; Cobut, Vincent[2]; Serfaty, Stéphane[1]

[1] Laboratory of Systems and Applications of Technologies applied to Information and Energy (SATIE), CNRS UMR 8029, ENS, CY Cergy Paris University, France; [2] Laboratory of Radiation and Matter in Astrophysical and Atmospheres (LERMA), CNRS UMR 8112, CY Cergy Paris University, France.

*Guyader, Natacha, 5 mail Gay Lussac, 95031 Neuville sur Oise, +33785345842, Natacha.Guyader@cyu.fr

**Introduction**.
New regulations as well as consumer preferences have prompted a shift in the cosmetic industry toward greener products based on natural raw materials and eco-friendly formulation processes [1], [2].
Regarding cosmetic emulsions, a wide array of possibilities is available for laboratories involved in cosmetic research. These products are obtained by emulsifying a fatty phase and an aqueous phase, and stabilized by surfactants [3]. They comprise multiple raw materials which differ both in their nature and origin. Each of them will have an impact on the emulsion properties, in particular on the chemicophysical parameters of the end-product, which can be measured and sometimes predicted. Moreover, the emulsion final characteristics can easily be modified by external parameters such as packaging, storage conditions, and transportation [4], [5].
Making appropriate choices to optimize the stability of a particular emulsion represents a challenge with mid-to-long-term consequences and is therefore of paramount importance.
The design of a new emulsion using a defined set of raw materials requires preliminary systemic studies, leading to emulsion cartography, such as HLD-WOR (Hydrophylic Lipophilic Deviation – Water-to-Oil Ratio) and Winsor type diagram, based on the number of phases and the emulsion type (Oil-in-Water (O/W) or Water-in-Oil emulsions (W/O)) [6]. This work intensive process would therefore benefit from a quick and easy screening method of the different parameters (intrinsic and extrinsic) of the emulsion [7], [8].
Two method paths are available to design stable emulsions with limited resources at hand. These cost-and-time-efficient approaches called experimental research methodologies explore factor impact (environmental ones such as temperature, or inherent like composition) before optimizing formulations [9]. The first path consists in a multidimensional analysis based on a dichotomic approach which requires to change parameters one by one (trial-and-error approach) [10], [11]. This method is commonly known to be quite ineffective to accurately correlate parameters. The second path, based on Design of Experiments (DOE), requires only a restricted number of experiments, followed by an extrapolation of missing data points. Unfortunately, it usually requires dedicated and expensive software and specific mathematical skills, especially for the change of multiple factors (more than two) [12]. In fact, numerous design of experiment models and algorithms are existing depending on the application (prediction or parameters study) [1], [5], [10], [11], [13]–[17]. These software are mostly based on pre-analysis, called screening design, to determine the factor levels. Then an optimal design can complete the analysis in order to optimize particular parameters [18], [19]. Some DOE software are also only specialized in a certain field of application, like the Experimental Design Assistant (EDA) for animal research [20]. The design of these experiments require time and precision to avoid mistakes due to a wrong factor levels determination [21]–[23]. These tests can be completed by Deep Learning or Neural Networks to optimize multiparametric issues. These processes involve time, specific tools and skills and it is quite complex [24].
Our work aims to provide a new screening method (FSM: Fast Screening Method) by defining the stability zones, especially for O/W emulsions, depending on the proportions of each of the raw materials. This method can also be used to identify the parameters influencing the end-product-state with only a few measurements. It provides a resource-effective way to plan experiments based on a random even distribution of emulsion phase ratios (Surfactants/Oil/Water: SOW) in order to identify domains of O/W emulsion stability.
To show the method effectiveness, simple O/W emulsions were formulated with different surfactants and oils. It is thus possible and easy to compare the impact of a raw material or a process change.

Additionally, principal component analysis (PCA) is used to complete the observations and link the stability domains to organoleptic properties of the emulsions. This method is voluntarily simple to be usable by small companies with limited resources as well as industrial groups.

We will first describe our method to choose random combinations of emulsion phase components in order to create simple emulsions; then, we will present how to determine experimentally the fine stability areas of emulsions of four sets of raw materials; finally, we will describe the links between the stability areas previously defined and the organoleptic properties of the emulsions using Principal Component Analysis.

The present work is focusing on the effects of the emulsion composition and phase ratios, with two types of oil origin and three different surfactants. Furthermore, our methodology can easily be adapted to investigate the impact of other parameters, such as pH or temperatures.

**Materials and Methods**.

*Method for combination determination:*

The fast screening method (FSM) uses an evenly distributed scanning of the three-phases composition represented in a triangle chart, in which each side describes an emulsion phase proportion.

A reasonable minimum difference between two different phase proportion of 5% was chosen in order to overtake the uncertainties of manipulation. 171 combinations of Surfactants/Oil/Water (SOW) are then possible. To reduce the number of trials, a set of combinations is pseudo random determined, excluding the start-and-end values (0 and 100%) for the three phases.

The algorithm used to choose these combinations is available in Figure 1. An example for a 20% minimum difference is described in the Figure 2.

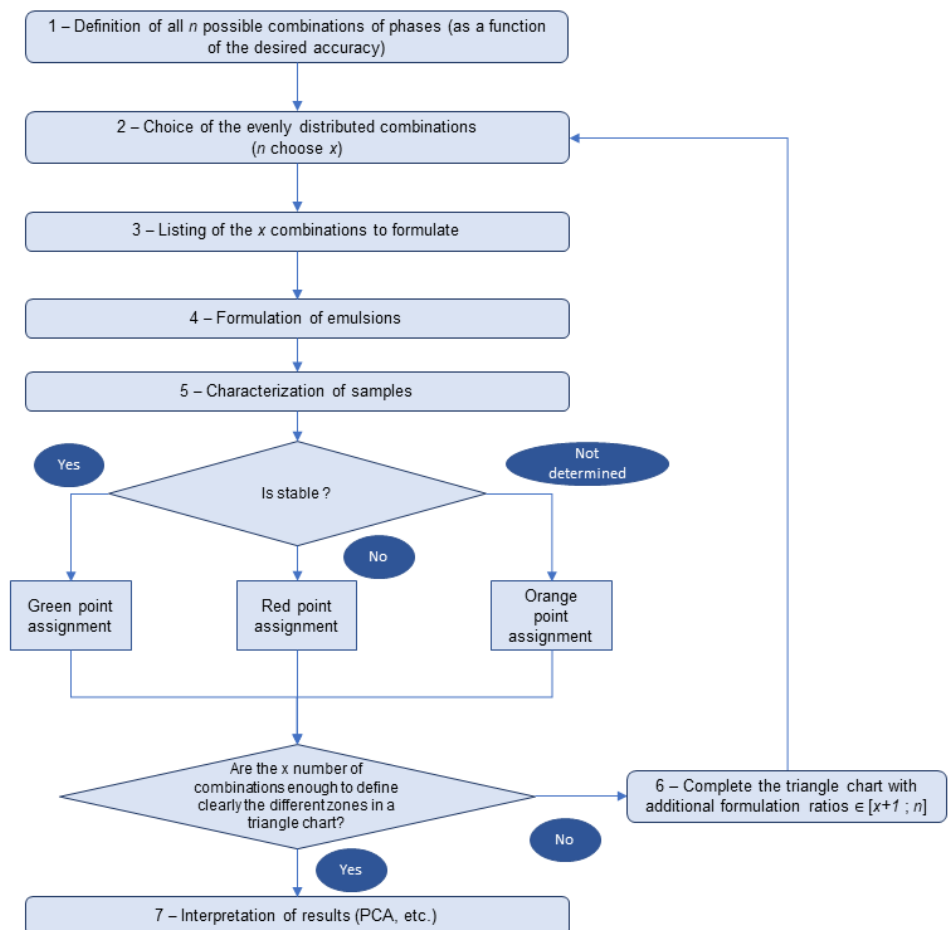

*Figure 1. Logic diagram flowchart for the method.*

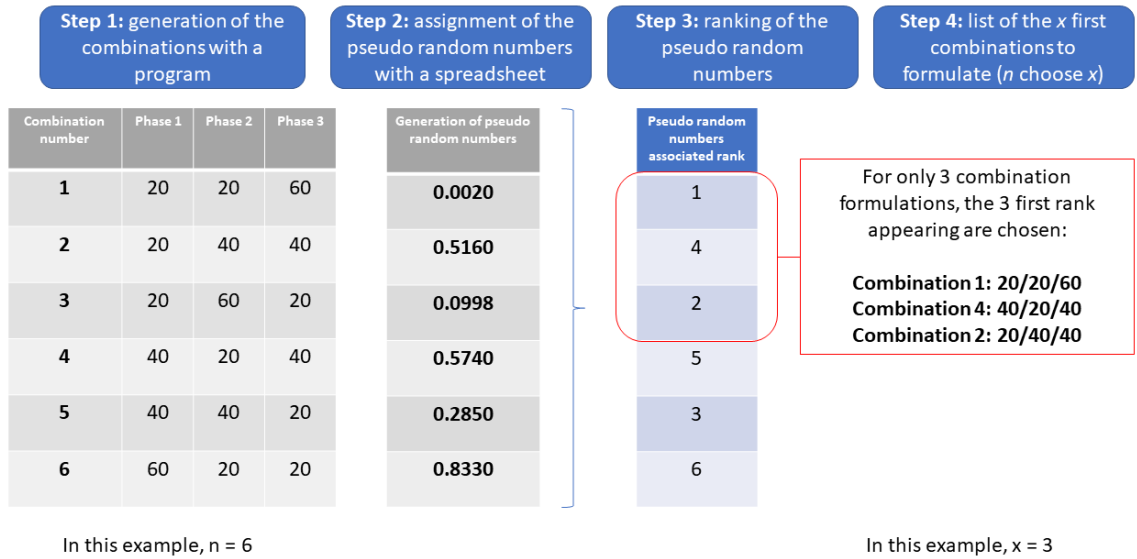

*Figure 2. Example of an experience planification for a 20% difference.*

In order to validate the method, a minimum of 30% of SOW combinations were tested. 49 combinations are defined in this study. An additional combination to qualify the repeatability is also realized for each set.

*Materials:*
For this study, simple emulsions are formulated with a couple of surfactants mixed with oil and gelified water. It voluntarily excludes additional raw materials such as conservatives or colouring. A list of the different used raw materials is presented in Table 1.

| Raw material category | INCI name | Origin | Colour | State | HLB | Density |
|---|---|---|---|---|---|---|
| Water | Demineralized water | Natural | Transparent | Liquid | / | 0.997 |
| Gelling agent | Carbomer | Chemical | White | Solid | / | / |
| Oil | Isopropyl palmitate (IPP) | Chemical | Transparent | Liquid | 11.5 | 0.851 |
| | Sunflower Seed Oil (SO) | Natural | Light yellow | Liquid | 10 | 0.919 |
| Surfactant | Polysorbate 80 (P80) | Chemical | Brown | Viscous liquid | 15 | 1.074 |
| | Oleth-2 | Chemical | Light yellow | Liquid | 4.0 | 0.912 |
| | Glyceryl stearate (GMS) | Chemical | White | Solid | 3.8 | 0.920 |

*Table 1. Raw material liste.*

*Method of formulation:*
Four triangle charts are designed from the three phases composing the emulsion (surfactant, oil and water phases: SOW), according to the following combinations: A: P80 + GMS / IPP / Water + Carbomer, B: P80 + Oleth-2 / IPP / Water + Carbomer, C: P80 + GMS / SO / Water + Carbomer and D: P80 + Oleth-2 / SO / Water + Carbomer. The raw materials used in the continuous/aqueous phase are demineralized water with a carbomer (0,3%). For the dispersed phase, different oils were used such as a natural oil (sunflower oil) or a chemical one (isopropyl palmitate). For each combination, two surfactants (Polysorbate 80 and Oleth-2 or Glyceryl Monostearate) have been mixed so that their combined mixing HLB matches the HLB required for the oil, as indicated in Table 1. 10 g of emulsion is formulated in test tubes for each trial, as follow. Both surfactants and aqueous phase are heated up to 75°C before being added to the oil at room temperature. This mixing is emulsified with a vortex 1 minute. Once back at room temperature, the pH is then adjusted at 5.5, by adding the 10% triethanolamine (TEA) solution (TEA/Water = 2.26) to thicken the emulsion thanks to the carbomer. Each sample is stored at 25°C for 24 ± 2 hours before the organoleptic analyses.

*Method of characterization:*
Five organoleptic descriptors were used to characterize the emulsions: the number, the thickening property, the opacity, the colour of phases and the presence of granules. Each of these factors is linked to a specific rating score. The number of phases goes from 1 to 3, based on the possible observations according to Winsor III diagram properties of the emulsion. The range of value are defined from 0 to 5 for the thickness and the opacity. For the thickening property 0 value corresponds to a solid state while 5 corresponds to the liquid state. The opacity range is defined linearly from opaque (0) to transparent (5). Three colours can be observed with these raw materials: white, yellow and transparent. The presence of granules on the inside of the emulsion is described with a Boolean description: absence or presence.

**Results**.
50 samples per set are achieved: 49 trials of formulation and a 50$^{th}$ one randomly duplicated to verify the reproducibility of the process. The repeatability of a set was also verified by changing the operator or the raw material batch. The results for each set are detailed in Figure 3.

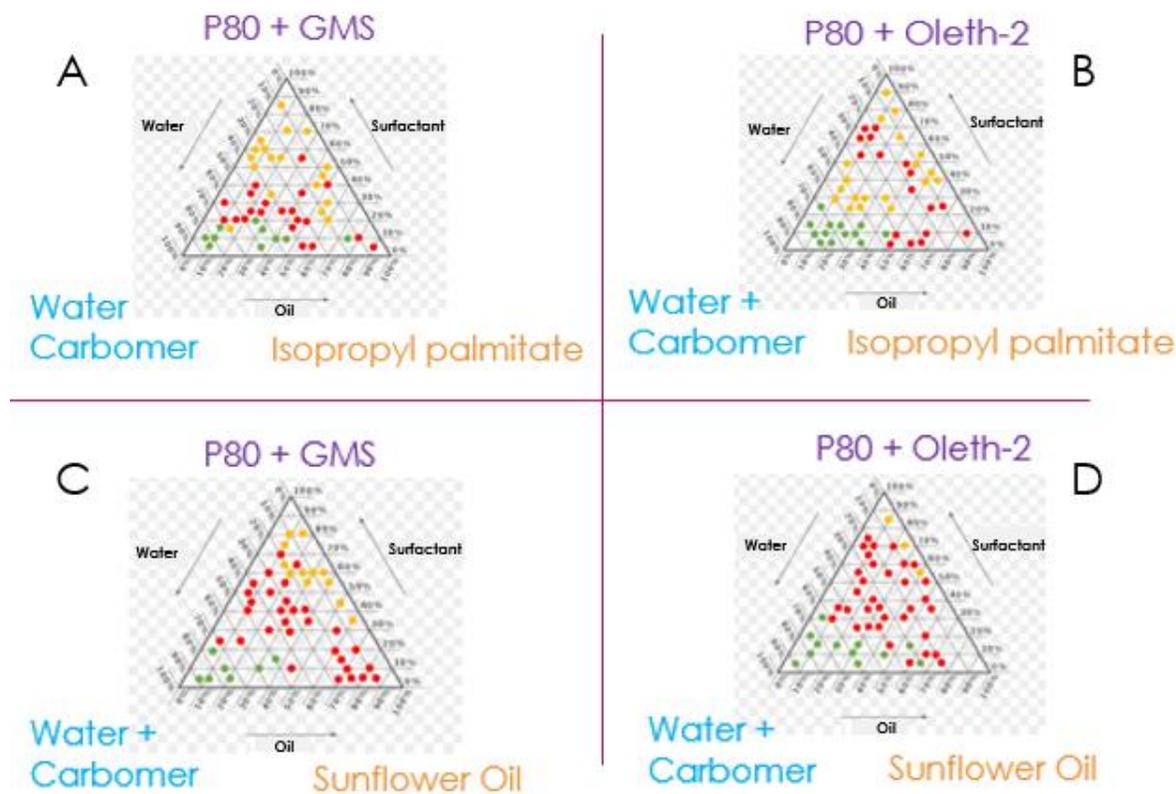

*Figure 3. The triangle chart realized with different SOW ratios. These charts describe the trial behaviour after 24h at 25°C and pH = 5.5. The spot colour is only based on the number of phases: green for O/W emulsions; red for unstable emulsions (several phase behaviour) and orange for stable emulsions except O/W. A: P80 + GMS / IPP / Water + Carbomer, B: P80 + Oleth-2 / IPP / Water + Carbomer, C: P80 + GMS / SO / Water + Carbomer and D: P80 + Oleth-2 / SO / Water + Carbomer.*

The triangle chart aims to easily detect the stability zone. For instead, with the examples above, the stability zones are better defined when sunflower oil is used (C, D) than when isopropyl palmitate (A, B). However, there are more unstable combinations (in red) with sunflower oil (C, D) than with isopropyl palmitate (A, B). Whatever the couple of surfactants used, the nature of oil seems to have a strong impact on the stability, probably due to the mix of fatty acids present in the natural oil, which could modify the HLD of the oil [25].

In order to evaluate the robustness of the stability zone detected and the organoleptic descriptors impacted by the emulsion composition, a Principal Component Analysis (PCA) is run. This method gives access to the greatest influence quantity impacting on the stability for each raw material set.

Note that the considered O/W optimum emulsion in this work has the following characteristics: number of phases (one), granularity (none), opacity (opaque), colour (white) and thickening (creamy). Then 5 vectors are considered in our PCA for each organoleptic descriptor. The PCA for the Figure 4 and Figure 5 only present the impact of descriptors on stable combinations (emulsion with only one phase) while the whole combinations are represented in Figure 6 and Figure 7.

The Figure 4 describes the analysis for the set A.

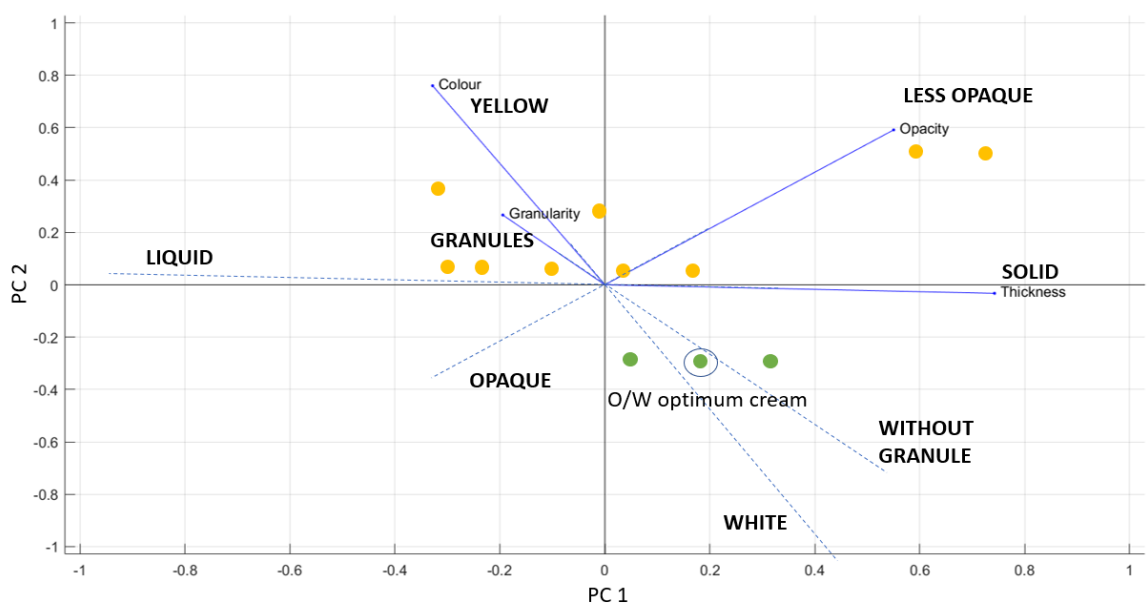

*Figure 4. PCA for the set A: P80 + GMS / IPP / Water + Carbomer. In green the O/W emulsions and in orange the stable emulsions except O/W.*

On this first analysis, the O/W stability zone is clearly defined as a function of the thickening parameter. Each state of stability can be then transposed in "three" horizontal lines. The position of these lines in the 2D PC plane evolve according to the colour, the opacity and the granularity. The bottom one corresponds to stable O/W emulsions (creamy), the upper ones correspond to the other stability states (gels, W/O emulsions, etc.). Similarly, the most affected descriptor which discriminates the O/W stability zone versus the other stability zone is the colour.

This representation can then objectivate the known impact of the components in emulsions. In our study, the position of the O/W emulsion creamy formulations are driven by the surfactant quantity. This effect is due to the initial solid state of one of the used surfactants. Increasing the quantity of solid surfactant leads to a thicker emulsion.

Considering the other stable emulsions (in orange), the opacity is linked with the ratio Surfactant/Aqueous phase, while the granularity depends on the oil quantity; if there is a lack of oil, the surfactant can aggregate in granules [26].

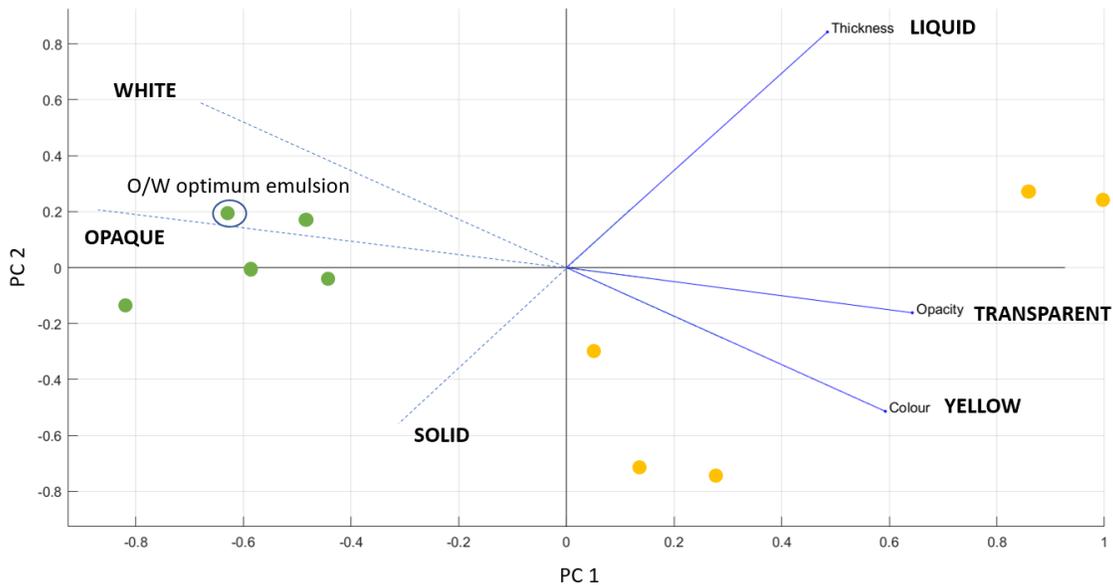

*Figure 5. PCA for the set B: P80 + Oleth-2 / IPP / Water + Carbomer mixing combinations. In green the O/W emulsions and in orange the stable emulsions except O/W.*

As shown in Figure 5 for the set B, the change of the solid surfactant for a liquid one avoids the granularity effect.
The behaviour of the emulsion and the zone of stability stays uniform.
However, if sunflower oil containing a complex fatty acid mix is selected, these zones risk to be reduced or separated, as shown in Figure 6 and Figure 7.

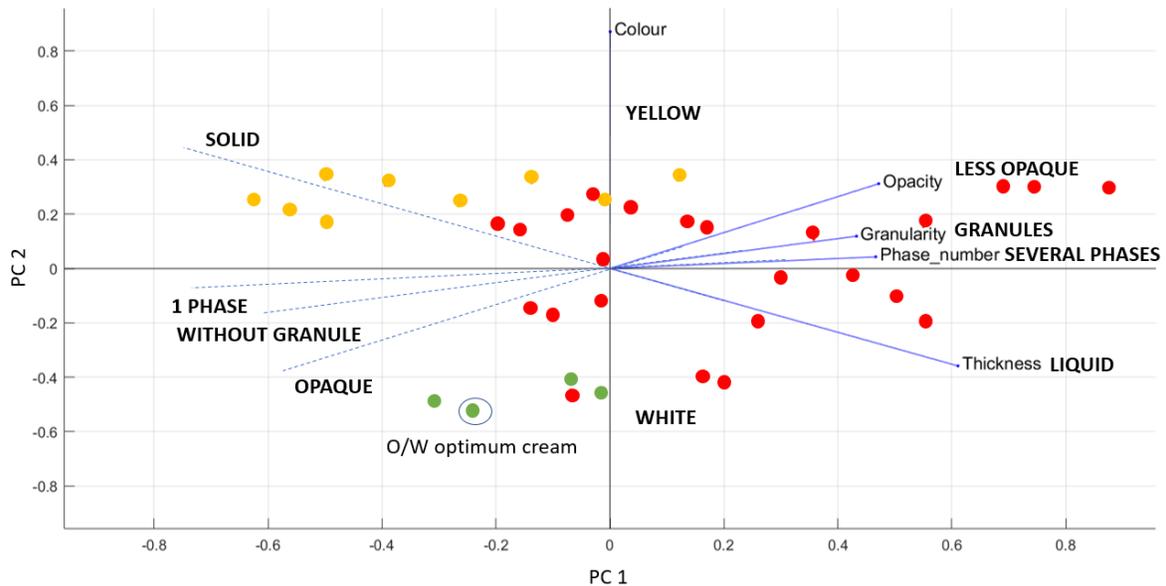

*Figure 6. PCA for the set C: P80 + GMS / SO / Water + Carbomer mixing combinations. Note that the unstable points were kept to highlight the stable/unstable crosslinked area. In green the O/W emulsions; in red the unstable emulsions (several phase behaviour) and in orange the stable emulsions except O/W.*

Indeed, in Figure 6, it can be noticed that two distinguished zones of stability are possible. To formulate an O/W emulsion it is quite difficult to be sure of the final state because of the presence of an unstable point in between the stable ones. The appearance of this stable/unstable crosslinked area suggest a possible impact during the scale up.

The use of a liquid surfactant can be then relevant as shown in Figure 7.

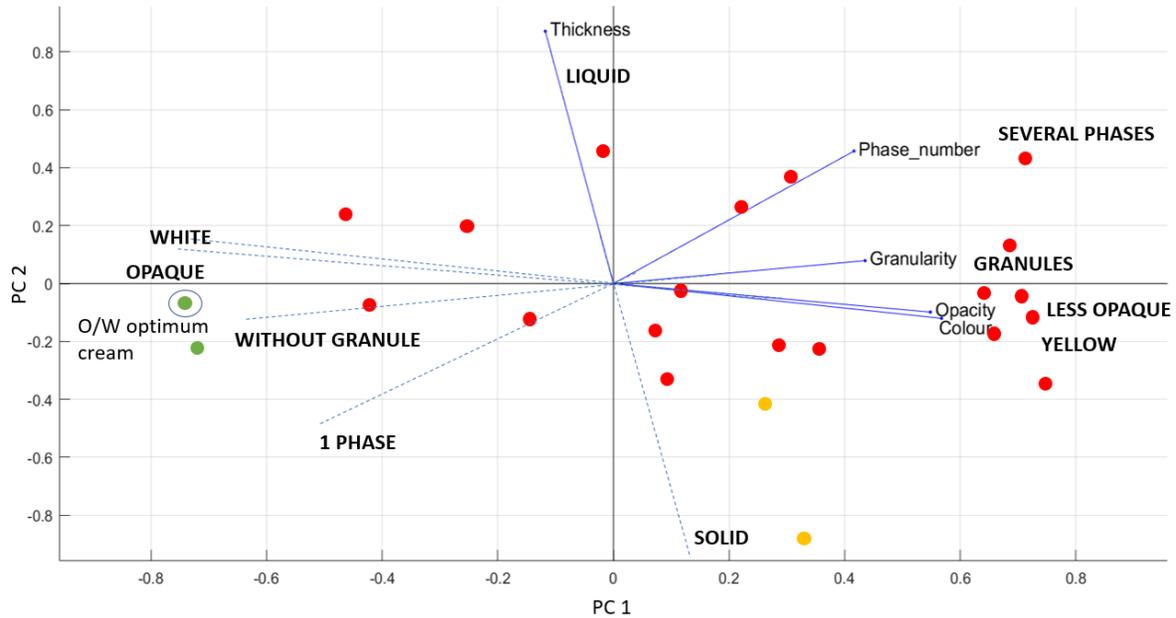

*Figure 7. PCA for the set D: P80 + Oleth-2 / SO / Water + Carbomer mixing combinations. In green the O/W emulsions; in red the unstable emulsions (several phase behaviour) and in orange the stable emulsions except O/W.*

Two important points can be highlighted in this PCA. On one hand, an O/W stable area is distinguishable in both Figure 3 (D) and Figure 7. On the other hand, several combinations allow to obtain two O/W optimum cream for which the organoleptic observations are almost similar. Only the thickness is lightly evolving.

These comparisons between triangle charts and PCA allow us to get complementary information on the impact of both composition (raw material nature and proportion) and organoleptic characteristics on stable or unstable emulsions. For these four triangle charts, it is possible to define the different zones of phase-behaviour. These results are considered reliable after repeating each in the different conditions quoted before.

In order to evaluate the time evolution of stability, a comparison of the triangle charts A and B from Figure 3 with a non-emulsified triangle (phases put in contact and not emulsified, for 180 days at 25°C), in concordance with a Winsor protocol, is presented in Figure 8 [27].
Here, the unstable (red points), stable O/W (green points) and other stabilities emulsions (orange points) are characterized the same way as before.
As shown in Figure 8, one can see that the observations after a long duration without emulsification are comparable with our FSM (Figure 3).

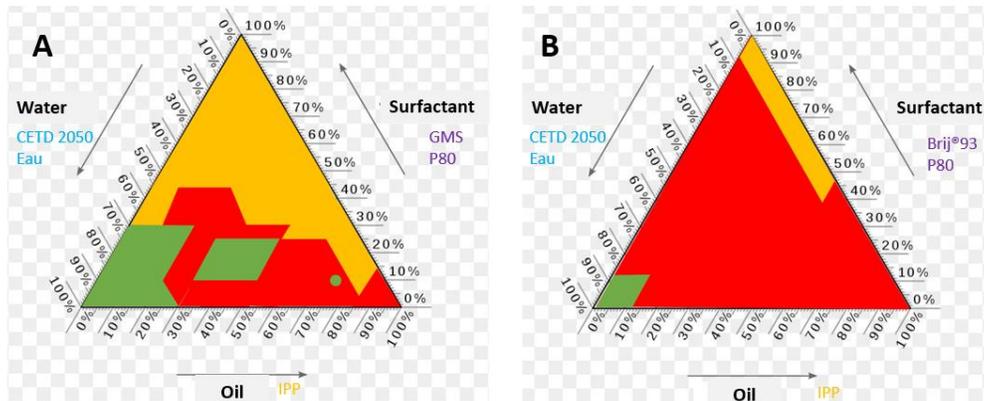

*Figure 8. The triangle chart realized with different SOW ratios. These charts describe the non-emulsified trial behaviour (raw materials put in contact) after 6 months at 25°C. The spot colour is only based on the number of phases: green for O/W emulsions; red for unstable emulsions (several phase behaviour) and orange for stable emulsions except O/W. A: P80 + GMS / IPP / Water + Carbomer, B: P80 + Oleth-2 / IPP / Water + Carbomer.*

From the sets A and B, all stables O/W zones are identified by the FSM. The main difference is on the supposed W/O emulsion after emulsification (in orange). Some of them tend to be unstable after 6 months. Our previously exposed method allows a prediction of stability in only 24 hours.

**Discussion.**
The method presented in this paper voluntarily focuses on organoleptic properties to show its easy implementation. However, other characterizations techniques, such as optic measurements can be used to complete the observations. Other parameters such as the pH or the temperature can also be changed in order to realize a complete pyramid or only another diagram cut (γ, etc.) [28]. This method was used according to the Δ cut diagram at constant storage temperature but it can also be done at different temperatures to create a 3D map of formulation. This FSM allow to screen quickly the map of formulation before entering into further analysis on the most interesting emulsions.

This method aims to determine preventively the zone of interest for an accurate longer-term study. Using appropriate characterization technics, the PCA can also help to define the specific influence quantity.

As shown in the PCA, the opacity, the thickness and the granularity play a big role in the characterization of emulsions. A low opacity is usually linked to a better stability, such as the micro or nano-emulsions but in this study, the droplet size is around a few microns.

On the same way, the thickness of cold cream compared to an everyday light cream are pretty different but is seems not possible to play with a large range of thickening values, unless going any further in the unstable zone. Finally, even in the emulsions composed of fully liquid components, a granularity can be observed. This is the beginning of some destabilization process mechanisms which at a certain stage can be observed with a rookie eye. The use of natural oil, such as sunflower oil, can lead to different stable zones on the PCA (Figure 6). However, it means that the emulsions don't have the same organoleptic properties. It represents a risk at the industrial scale to switch from a smooth emulsion to a grainy one, especially for the customer expectations.

The same method can be used to redefine more clearly the boundary between two zones.

One combination of each zone could also be studied at a higher scale with a tank to evaluate and characterize destabilization mechanisms with multiscale measuring instruments. Using an effective planning of experiments constitutes a research asset by reducing the number of experiments and resources (time and raw materials) but also allowing to predict the different zones of stability to further study O/W emulsion instability. If the method allows to find several stable zones with the same organoleptic descriptors, it is also possible to focus on a working zone for the emulsion that will be industrialized according to the raw material prices.

The low quantity samples also allow to realize complementary tests with multi-scale characterization techniques such as low frequency rheology or pH and conductivity measurement.

**Conclusion**.
In this study, a complete understanding of the global set up was essential to understand what was at stake with the outputs, upstream the use of a DOE software. Thanks to an easy methodology using pseudo-random numbers and PCA, it is possible to use fast predictive methods to study the long-term stability of new emulsions. This paper aims to highlight the ability of the FSM.

**Conflict of Interest Statement**.
NONE

**References**.